\documentclass[lettersize,journal]{IEEEtran}
\usepackage{amsmath,amsfonts}
\usepackage{algorithmic}
\usepackage{array}
\usepackage[caption=false,font=normalsize,labelfont=sf,textfont=sf]{subfig}
\usepackage{textcomp}
\usepackage{stfloats}
\usepackage{url}
\usepackage{verbatim}
\usepackage{graphicx}
\def\BibTeX{{\rm B\kern-.05em{\sc i\kern-.025em b}\kern-.08em
    T\kern-.1667em\lower.7ex\hbox{E}\kern-.125emX}}
\usepackage{balance}
\usepackage{amssymb}
\usepackage{cite}
\usepackage{amsmath,amssymb,amsthm,mathtools}
\usepackage{bm}
\usepackage{booktabs}
\usepackage{mathrsfs}
\usepackage{tikz}

\usetikzlibrary{arrows.meta,positioning,calc,fit,decorations.pathreplacing}

\newcommand{\E}{\mathbb{E}}
\newcommand{\Prob}{\mathbb{P}}
\newcommand{\ind}{\mathbf{1}}
\newcommand{\dd}{\,\mathrm{d}}
\newcommand{\disk}{\mathbb{D}}
\newcommand{\sphere}{\mathbb{S}^{2}}
\newcommand{\capset}{\mathcal{C}}
\newcommand{\voidfun}{\mathcal{V}}

\newtheorem{theorem}{Theorem}
\newtheorem{proposition}{Proposition}

\newtheorem{corollary}{Corollary}
\theoremstyle{definition}
\newtheorem{definition}{Definition}
\theoremstyle{remark}

\begin{document}
\title{A User-Centric Handover Analytical Model for LEO Satellite Networks }
\author{Bin Liu,~Haifeng Hu,~Lin Chen~\IEEEmembership{Member,~IEEE},~and Hui Zhang~\IEEEmembership{Member,~IEEE}
	\thanks{This work was supported in part by Enterprise Innovation and Development Joint Fund of the National Natural Science Foundation of China under Grant No.U24B20187, and in part by National Natural Science Foundation of China under Grant No.62371245. (\emph{Corresponding author: Haifeng Hu})}
	\thanks{Bin Liu, Haifeng Hu, and Hui Zhang are with the College of Telecommunications and Information Engineering, Nanjing University of Posts and Telecommunications, Nanjing 210003, China. (e-mail: liubin0430@njupt.edu.cn, huhf@njupt.edu.cn, 20080017@njupt.edu.cn).
	}
	\thanks{Lin Chen is with Engineering Research Centre of Applied Technology on Machine Translation and Artificial Intelligence, Macao Polytechnic University, R. de Luís Gonzaga Gomes, Macao, China. Post Code: 999078 (e-mail: lchen@mpu.edu.mo)}
	\thanks{}
}

\markboth{Journal of \LaTeX\ Class Files,~Vol.~18, No.~9, September~2026}%
{How to Use the IEEEtran \LaTeX \ Templates}

\maketitle

\begin{abstract}
The continuous large-scale deployment of Low Earth Orbit (LEO) satellite networks has motivated the study of dynamic network performance, for which the handover intensity, or handover rate, is a key metric characterizing how frequently the serving satellite changes. However, existing studies largely rely on simulation or static approximations and lack an exact, geometrically interpretable characterization of handover rate in dynamic LEO networks. 
To quantify how constellation parameters affect handover rate, this paper develops a trajectory-based analytical framework to quantify handover performance from the viewpoint of a typical ground user. 
The constellation consists of circular orbits at a common altitude with a common angular speed, Poisson orbital planes, and Poisson satellite phases on each orbit. Starting from the three-dimensional Earth–satellite geometry, we establish an exact transformation that maps each moving satellite to a fixed point in a unit disk, referred to as the projection space. The satellite association decision then reduces to maximizing a cosine distance functional that varies with the rotation direction over time, so that only the transformed node with the maximum closeness can serve the user. 
We further show that the transformed constellation remains a radial Cox process, whose probability generating functional and reduced Palm distribution are exploited to derive dynamic network performance. A rigorously proved theorem provides the general handover intensity, and separate same-orbit and different-orbit handover intensities are derived. An illustrative example demonstrates the effectiveness of the framework, and extensive simulations validate the analysis. 
By presenting fundamental handover intensity as a key parameter, this analytical framework allows us to assess more statistical properties of LEO satellite networks. Thus, It can be used as a system-level design tool to optimize LEO constellations. 

\end{abstract}


\section{Introduction}
\IEEEPARstart{L}{ow}-Earth-orbit (LEO) satellite constellations, such as Starlink and OneWeb, are increasingly deployed to provide global broadband access. Their dynamic nature, however, poses fundamental challenges for service continuity over finite ground user's local horizon or field-of-view (FoV). Specifically, unlike geostationary satellites, LEO satellites move rapidly relative to ground users, which means that both the serving link and the interference environment change relatively quickly over time.
Furthermore, the massive deployment of LEO satellites leads to increasingly frequent link switches among neighboring nodes. As a satellite moves, another satellite may become closer and cause the serving identity to change rapidly. Immediately after that link switching, the new satellite may remain the best visible satellite for a long interval, or it may lose that role almost immediately. These complicated network dynamics not only burden system signaling cost but also significantly impact link stability.

Hence, satellite constellations must be carefully designed to provide both high-speed, broadband and reliable service to ground users. In general, providing such a service must account for two coupled factors: (i) time-varying satellite positions and identities; and (ii) physical-layer outage constraints based on simple line-of-sight (LoS) visibility. These two factors collectively render the service continuity problem intrinsically finite, i.e., limited to the ground horizon in the conventional slot-wise sense. To address this issue, it is imperative to investigate dynamic satellite network performance under different constellation deployments.

Unfortunately, network dynamics cannot be captured by a snapshot coverage probability, since it averages over a randomly selected time instant and is blind to identity changes. What is required is a duration or link statistic that resolves the service lifetime of a typical satellite.
The most conventional approach is to rely on large-scale Monte Carlo simulation to evaluate average link performance and duration. This is feasible for small-scale satellite networks but becomes computationally demanding and questionable for modern LEO networks with a large number of satellites. In addition, such exhaustive simulations lack an exact, geometrically interpretable characterization of link switching, or \emph{handover}, in dynamic LEO networks.

Since the spatial distribution of satellites directly determines the link state and performance, stochastic geometry, which has a long history in modeling wireless networks, has recently become a powerful tool for building analytical models and obtaining tractable statistical performance for satellite networks. This approach paves the way for studying the dynamic performance of large-scale satellite networks.

\subsection{Prior Work}
\subsubsection{Basic Stochastic Geometry Modeling}
Stochastic geometry has been widely used to represent satellite locations and to derive snapshot performance measures, including coverage and association probabilities.  Early work \cite{Homssi2021,Hourani2021} attempted to employ it to model dense satellite locations for uplink and downlink communications, respectively. 
Okati \emph{et. al.} \cite{Okati2020} adopted a uniform binomial point process (BPP) to model LEO satellite systems as a replacement for Walker constellations, and showed that stochastic constellation geometry accurately captures satellite locations. This BPP model was later extended to a non-homogeneous Poisson point process (PPP) in \cite{Okati2022}, where the intensity is adjusted according to the actual satellite distribution. In the follow-up work \cite{Okati2026}, stochastic geometry was applied to analyze the coexistence of terrestrial and non-terrestrial networks in terms of coverage probability and average data rate. In \cite{Chae2023}, Chae \emph{et. al.} modeled satellite networks as a PPP to study directional beamforming gain, and showed that the PPP model yield a lower bound on performance compared with deterministic modeling.

Compared with BPP/PPP models, Choi \emph{et. al.} \cite{choi2024multi,choi2025cox} advocated the classical Cox point process to capture both the dependence among satellites in the same orbit and the independence across different orbits. Their results validate that the Cox point process approximates the signal-to-interference ratio (SIR) of Starlink remarkably well. Wang \emph{et. al.} \cite{Wang2025} summarized typical spherical stochastic geometry models, such as BPP, PPP, Cox, and dual-stochastic BPP, used for different constellations in recent years, and pointed out several research directions. In their follow-up work \cite{Ru2026}, a leader-follower satellite architecture was proposed to improve performance and reduce interference. Based on stochastic geometry, it was shown that the communication quality achieved by increasing the number of followers highly depends on reliable links with users.

Shang \emph{et. al.} \cite{Shang2026} leveraged stochastic geometry to investigate the statistical performance of multi-satellite cooperative communications. Besides conventional coverage probability and average data rate, they proposed several new metrics, such as non-handover probability, handover success rate, and multi-satellite spectral efficiency/cooperation gain. Unfortunately, how to derive these metrics was not discussed. In \cite{Li2026}, Li \emph{et. al.} introduced a time-delay outage probability metric to investigate the performance of multi-tier heterogeneous satellite networks, in which the service duration is incorporated into the coverage probability. However, the time required to traverse beyond the dome region of a serving satellite is not equivalent to the average service duration. Homssi \emph{et. al.} \cite{Homssi2025} developed an stochastic geometry-based theoretical model for the coverage probability of various diversity modes in multi-shell deployments of mega satellite constellations, and demonstrated the benefits of cross-tier cooperation. Lim \emph{et. al.} in \cite{Lim2024} analyzed interference effect between satellite and terrestrial networks. In addition, Liu \emph{et. al.} \cite{Liu2026} considered a three-dimensional (3-D) blockage model for satellite networks to capture the spatial correlation among serving and interference links under a PPP-based satellite location distribution. However, the time correlation induced by satellite motion was not addressed.

\subsubsection{Handover Performance based on  Stochastic Geometry}
Around 2021, A. Al-Hourani \emph{et. al.} \cite{Al-Hourani2021} studied the session duration when a satellite passes through a user's field-of-view.
Handover performance for drone cellular networks was investigated in \cite{Banagar2020}. However, how it can be adapted to satellite networks was not addressed.
In order to study the handover rate in space-ground integrated network, Zhou \emph{et. al.} \cite{Zhou2023} applied the a modified particle swarm optimization method to study the classical sum-rate problem where only user mobility was considered. 
A recent preprint paper \cite{Baccelli2026} written by Baccelli \emph{et. al.} studied the moving base station effect in dynamic wireless networks. To simplify the analysis, however, the base stations are restricted to a two-dimensional (2-D) plane. Consequently, the periodic satellite motion along orbits in 3-D space cannot be characterized by this 2-D model.
We note that a recent paper \cite{Homssi2026} addressed the handover rate in satellite networks. In that paper, the pass arc length of a satellite inside a typical user's field-of-view was first derived assuming random trajectory orientation; then, expected handover duration (e.g., persistent and fast handovers) were derived from the pass arc length distribution. Unfortunately, this method overlooks the satellite location distribution and cannot capture handovers caused by range variations of nearby satellites.

In summary, although the above works provide useful network-level performance characterizations, they either conducted only snapshot analyses or omitted handover due to the satellite location variations. 
A snapshot probability does not specify how long a typical serving satellite will remain selected, or equivalently, how frequently the service links switch. 
The omission of satellite location variations overestimates or underestimates link switches and performance as a result.
Hence, a dynamic satellite network must  account for changes of the serving satellite, often through handover counts, rates, or service duration. 
This is because the satellite observed at an arbitrary time is not distributed like the satellites selected at a handover, whose selection is weighted by relative speeds and locations, and may involve one or two parent orbits.
Therefore, a tractable yet realistic handover analytical model is highly desirable for dynamic satellite networks, especially for large-scale LEO satellite networks. Such a model should not only provide insight into statistical handover performance, but also help understand when and how link switches occur for a given satellite network. This paper aims to fill this gap.

\subsection{Challenges and Contributions}
Several challenges arise in analyzing handover performance in satellite networks.  \textbf{1) Handover conditioning and selection bias.} A handover is jointly induced by satellite motion, orbital geometry, and the association rule. It is therefore governed by the Palm distribution conditioned on a switching event rather than by the stationary distribution. This conditioning introduces a selection bias: only satellites of the same range that are more promising candidates can become the switching target.
\textbf{2) Shared orbit dependence and intractable integrals.} Satellites are constrained to move along circular orbits and are therefore strongly dependent. Classical independent BPP/PPP models fail to capture this dependence, whereas Cox point process models preserve it but substantially complicate the derivation of handover intensity, since the intensity must be averaged under the handover Palm measure. The resulting integrals are geometrically non-trivial, and closed-form expressions are rarely available.
\textbf{3) Periodic orbital motion in 3-D satellite networks.} Unlike 2-D models, where node motion is typically unstructured and translation invariant, LEO satellites move along circular orbits with a common angular speed, so that visibility, association, and handover inherit a periodic phase structure. This periodicity effects make a satellite may re-enter the visibility region after one orbital period and the handover statistics depend on the orbital phase rather than on time-translation invariance alone. Consequently, the handover process is neither a simple time stationary point process, nor a direct analogue of its planar counterpart. A tractable 3-D model must therefore represent the orbital phase explicitly and account for the periodic coupling between satellite motion and the association rule.

These challenges explain the reasons why most existing stochastic geometry-based work were confined to snapshot analyses. In this paper, we consider a typical ground user and a constellation of LEO circular satellite orbits at a common altitude and with a common angular speed.
The user direction is fixed in an inertial frame, and the user is associated with the nearest visible satellite. Relative to the satellite motion, the Earth's rotation is neglected. Under these assumptions, the association level of each satellite is a cosine distance with the same angular frequency but different initial phases. Hence, using sine and cosine basis vectors, the physical satellite positions are transformed to fixed points in a 2-D projection space. Time then appears only through a rotating query direction, and the serving satellite is the point that maximizes the corresponding cosine distance. This representation provides a geometric description of the serving sequence, rather than only a snapshot of it. It also preserves the dependence between satellites on the same orbit.  In projection space, conditional on the random orbit radii, the transformed satellites still form Poisson processes on orbit-specific circles, while their unconditional superposition is a radial Cox process. 
Finally, the handover event reduces continuous-time handover competition to the absence of competitors from a finite union of disk caps. Classical tools from reduced Palm theory \cite{Chiu2013,Baccelli2020Random} is utilized to derive the handover rate or intensity in combination with void competitor probability conditional on two satellites of same cosine distance.

To the best of our knowledge, this is the first analytical framework for studying handover performance in LEO satellite networks based on Cox process. This paper makes the following contributions:
\begin{enumerate}
	\item Starting from the 3-D Earth-user-satellite geometry, we devise an exact transformation that maps moving satellites in different orbits to fixed points in a projection space.
	
	\item We show that the nearest-visible-satellite association is represented as the maximization of a rotating linear functional, which connects the serving sequence to the transformed point set in the projection space.
	
	\item We characterize the transformed constellation as a radial Cox process and derive the background void probability and the tagged-orbit void probability. The resulting formulas retain the dependence induced by satellites sharing an orbit. We also express the coverage probability in terms of the void probability.
	
	\item Using reduced Palm conditioning, we derive separate expressions for the same-orbit and different-orbit handover intensities, together with the total handover intensity.
	
	\item We validate the handover intensity results with an illustrative example and conduct numerical simulations to evaluate the overall analytical framework.
\end{enumerate}

\subsection{Organization}
The remainder of this paper is organized as follows. Section~II describes the system model for satellite networks and the typical user location. Section~III presents the orbit geometry transformation used to construct the state point process in a projection space. Section~IV derives the same-orbit and different-orbit handover intensities from the obtained tagged-orbit void probability. Section~V provides an illustrative example and numerical results. Section~VI concludes the paper. The main used notations are listed in Table \ref{tab:notation}. In what follows, the terms \emph{handover intensity} and \emph{handover rate} are used interchangeably, as they refer to the same handover frequency.
\begin{table}[t]
	\caption{Main notations}
	\label{tab:notation}
	\centering
	\begin{tabular}{c|c}
\hline
		Symbol & Meaning\\
\hline
		$R_{\rm e},a,h$ & Earth radius, orbital radius, and altitude.\\
		$\bm u,U$ & Fixed user direction and location $U=R_{\rm e}\bm u$.\\
		$\bm n,\Pi$ & Oriented normal vector and physical orbit plane.\\
		$\bm e_{1},\bm e_{2}$ & Adapted orthonormal basis in orbit plane.\\
		$\theta_{\bm n, j}$ & Initial phase of satellite $j$ on orbit $\bm n$.\\
		$\omega,\varphi$ & Angular speed, current angle $\omega t$.\\
		$r_{\bm n}$ & Transformed orbit radius $\sqrt{1-(\bm u^{\mathsf T}{\bm n})^2}$.\\
		$\bm p=(A,B)$ & Static state point; closeness is $A\cos\varphi+B\sin\varphi$.\\
		$\bm w(\varphi)$ & Unit Rotational query vector $(\cos\varphi,\sin\varphi)^{\mathsf T}$.\\
		$c$ & Coverage threshold of minimum elevation.\\
		$\widetilde\Phi$ & State point process in the projection space.\\
		$\capset(\bm w,z)$ & Open cap $\{\bm p\in\disk:\bm p^{\mathsf T}\bm w>z\}$.\\
		$\voidfun(D)$ & Void probability of satellites in $D$.\\
		$z,b_i,b_j$ & HO closeness and outgoing/incoming coordinates.\\
		$\nu^{\mathrm s}_{\mathrm{HO}},\nu^{\mathrm d}_{\mathrm{HO}},\nu_{\mathrm{HO}}$ & Same-orbit, different-orbit, and total HO intensities.\\
		$\Xi,\Theta_{\bm n}$ & Orbit plane process and satellite phase point process.\\
		$\Lambda_{\rm orb},\Lambda_{\rm pha}$ & Orbit intensity measure and satellite intensity measure.\\
		$\lambda,\mu$ & Orbit intensity and satellite intensity per orbit.\\
		$\widetilde{\Lambda}_r,\widetilde{\Lambda}_{\Xi}$ & Intensity measure of transformed points in one/all orbits\\
\hline
	\end{tabular}
\end{table}

\section{System Model}
\subsection{Geometric Model}
\begin{figure}[!t]
	\centering
	\includegraphics[width=2.5in]{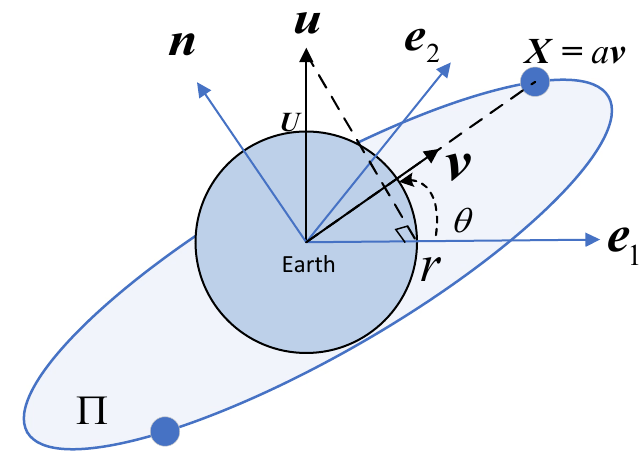}
	\caption{A schematic diagram illustrating the Earth, a typical user, the orbital plane, and the satellite phases.}
	\label{fig1}
\end{figure}

We consider the Earth in a $\mathbb{R}^3$ space, where $O$ denotes the center of the Earth. Let $R_{\rm e}=6371\,\mathrm{km}$ denote the Earth's average radius, and $\mathbb S^2=\{\bm n\in\mathbb R^3:\|\bm n\|=1\}$  denote the unit sphere. 
A typical user is located on the surface of the Earth and indicated by location vector $	\bm U=R_{\rm e}\bm u$, where $\bm u\in\mathbb S^2$ is the unit radial vector from the Earth's center to the user, as shown in Fig. \ref{fig1}. In order to simplify analysis, we treat $\bm U$ as fixed location and omit Earth rotation and user motion effect. Since the orbital model is isotropic, every fixed
choice of \(\bm u\) gives the same distribution. Hence, we set $\bm u=(0,0,1)^{\mathsf T}$ in our model.

We suppose all satellites move on ideal circular orbits of radius $a=R_{\rm e}+h$, 
where $h$ is the common satellite altitude. An orbit is characterized by an
oriented unit normal vector $\bm n\in\mathbb S^2$. We define that the orientation of $\bm n$ specifies the direction of orbital
motion through the right-hand rule. Thus, $\bm n$ and $-\bm n$ describe the same geometric plane but opposite directions of motion. With this orientation definition, a corresponding orbital plane is given by
\begin{equation}
	\Pi(\bm n)=
	\{\bm x\in\mathbb R^3:\bm n^{\mathsf T}\bm x=0\}.
\end{equation}

All satellites are assumed to have the same positive angular speed $\omega$, which satisfies the physical acceleration equation  $a\omega^2=\frac{G M_{\rm e}}{a^2}$, where $G$ is the gravitational constant, $a$ is the orbit radius, and \(M_{\rm e}\) is the Earth's mass. This gives $\omega=\sqrt{G M_{\rm e}/a^3}$. As a result, the orbital period is $2\pi/\omega$.

For each orbit plane normal vector $\bm n\in\mathbb S^2$, we define orbit parameter $	r(\bm n)=\sqrt{1-(\bm u^{\mathsf T}\bm n)^2}.$
When \(r(\bm n)>0\), we choose the in-plane orthonormal basis
\begin{align}
	\bm e_1(\bm n)
	&=
	\frac{\bm u-(\bm u^{\mathsf T}\bm n)\bm n}
	{r(\bm n)},
	\label{eq:e1}
	\\
	\bm e_2(\bm n)
	&=
	\bm n\times\bm e_1(\bm n),
	\label{eq:e2}
\end{align}
where $\times$ denotes vector product. These vectors lie in orbit plane $\Pi(\bm n)$ and satisfy relations among three directional basis $\bm e_1(\bm n)\times\bm e_2(\bm n)=\bm n$. With these basis definitions, we have the relations
\begin{equation}
	\bm u^{\mathsf T}\bm e_1(\bm n)=r(\bm n),
	\quad
	\bm u^{\mathsf T}\bm e_2(\bm n)=0.
\end{equation}
It should be stressed that $\bm e_1(\bm n)$ is the in-plane direction onto which the projection of the user direction $\bm u$ is the largest, as shown in Fig. \ref{fig1}. It also defines the
zero-phase direction, while \(\bm e_2(\bm n)\) determines the direction of increasing phase consistently with the orientation of $\bm n$.

With bases $\bm e_1(\bm n),\bm e_2(\bm n),\bm n$, a unit position direction on the orbit with initial phase $\theta$ can be expressed as
\begin{equation}
	\bm v(\bm n,\theta)
	=
	\cos\theta\,\bm e_1(\bm n)
	+
	\sin\theta\,\bm e_2(\bm n),
	\qquad 0\leq\theta<2\pi,
\end{equation}
and the corresponding physical satellite position can be simplified as $\bm X(\bm n,\theta)=a\bm v(\bm n,\theta).$
Particularly, $\bm u^{\mathsf T}\bm v(\bm n,\theta)=r(\bm n)\cos\theta.$ The relations among these orbit and satellite parameters are shown in Fig. \ref{fig1}.
When $r(\bm n)=0$, the user direction is parallel to the orbit normal vector and does not define a unique in-plane phase origin. This trivial case has probability zero in our 3-D orbit plane model.

Then, let $\{\theta_{\bm n,j}\}_j$ be the satellite initial phases relative to the reference time $t=0$ for a given orbit $\bm n$ . At time $t$, the unit position vector and the physical position of satellite $j$ are respectively written as
\begin{align}
	\bm v_{\bm n,j}(t)
	&=
	\cos(\theta_{\bm n,j}+\omega t)\,\bm e_1(\bm n)
	+
	\sin(\theta_{\bm n,j}+\omega t)\,\bm e_2(\bm n),
	\label{eq:unit-trajectory}
	\\
	\bm X_{\bm n,j}(t)
	&=
	a\bm v_{\bm n,j}(t).
	\label{eq:trajectory}
\end{align}
In what follows, we will use these equations to derive satellite positions in a projection space. 

In addition, since this paper focuses on handover among neighboring satellites as the ranges vary, all satellites are assumed to be equipped with omnidirectional antennas with the same unit antenna gain and transmission power. Large-scale path loss and small-scale fading effects are omitted for brevity. The pass duration of each satellite is affected only by a typical user's field-of-view and a predefined elevation threshold, as detailed in Section III.
\subsection{Cox Satellite Point Process}
We utilize the classical Cox point process \cite{choi2025cox,choi2024multi} model to characterize the constellation statistical distribution.  Let $\Xi=\{\bm n_1,\bm n_2,\bm n_3,...\}$ be the parent \emph{orbital plane process}, which is a Poisson point process on $\mathbb S^2$ 
with intensity measure
\begin{equation}
	\Lambda_{\rm orb}(\dd\bm n)
	=
	\frac{\lambda_{}}{4\pi}\dd\sigma(\bm n),
\end{equation}
where \(\dd\sigma(\bm n)\) is surface area measure on $\mathbb S^2$. The intensity $\lambda_{}$ is the average orbit number per unit area.
Conditional on the orbit $\Xi$ process, the satellites on each orbit form independent satellite \emph{phase point processes}. Specifically, for each $\bm n\in\Xi$, define $\Theta_{\bm n}$ as a Poisson point process on angular range $[0,2\pi)$ with intensity measure
\begin{equation}
	\Lambda_{\rm pha}(\dd\theta)
	=
	\frac{\mu}{2\pi}\dd\theta,
	\qquad 0\leq\theta<2\pi,
	\label{eq:phase-intensity}
\end{equation}
where \(\mu\) is the mean number of satellites per orbit. The phase points in collection $\Theta_{\bm n}$ are the satellite position phases at reference time $t=0$.
All satellites lie on the sphere $\mathcal S_a=\{\bm x\in\mathbb R^3:\|\bm x\|=a\}.$
Then, satellite point process at the reference time is given by
\begin{equation}
	\Phi_{\rm sat}
	=
	\sum_{\bm n\in\Xi}
	\sum_{\theta\in\Theta_{\bm n}}
	\delta_{\bm X(\bm n,\theta)},
	\label{eq:satellite-point-process}
\end{equation}
where satellite physical location $\bm X(\bm n,\theta)$ is identified by orbit $\bm n$ and initial phase $\theta$. Since each phase point $\theta \in \Theta_{\bm n}$ can be independently displaced to position on $\mathcal S_a$, it follows that, conditional on \(\Xi\), \(\Phi_{\rm sat}\) is a Poisson point process on \(\mathcal S_a\) with the following random intensity measure, i. e., for a measurable set $A$,
\begin{align}
&\Lambda_{\Xi}(A)=\mathbb{E}\bigl[\#\{\text{points in }A\}\mid \Xi\bigr]\notag\\
&=
	\frac{\mu}{2\pi}
	\sum_{\bm n\in\Xi}
	\int_0^{2\pi}
	\ind\{\bm X(\bm n,\theta)\in A\}\,\dd\theta,
	\quad
	A\subseteq\mathcal S_a\ .
	\label{eq:conditional-satellite-intensity}
\end{align}
Its mean total number of satellites is $\mathbb E\!\left[\Phi_{\rm sat}\right]=\lambda\mu.$

\begin{figure}[!t]
	\centering
	\includegraphics[width=2.5in]{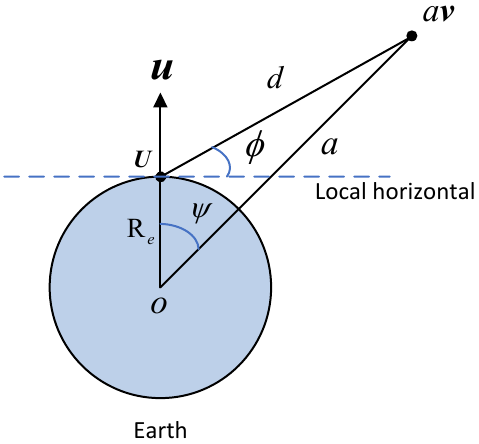}
	\caption{The geometric relation of the closeness and visibility threshold. The Earth-centered angle is $\psi$, whereas the elevation angle $\phi$ is the one measured at the user.}
	\label{fig2}
\end{figure}

\section{Orbit Geometry Transformation}
\label{sec:transformation}
\subsection{Closeness Metric}
In real satellite network deployment, the serving node generally is the closest satellite with the minimal geocentric angle with respect to the ground user. Motivated by this, we first define a user-centric proximity metric. Specifically, for satellite $j$ on the orbit with normal vector $\bm n$, define
\begin{equation}
	s_{\bm n,j}(t)
	:=
	\bm u^{\mathsf T}\bm v_{\bm n,j}(t).
	\label{eq:closeness}
\end{equation}
Obviously, this is the cosine distance of the Earth-centered angle between the typical user
direction and the satellite direction, as shown in Fig. \ref{fig2}. The larger this distance, the closer the satellite is to the typical user. The user-satellite distance is given by
\begin{align}
	d_{\bm n,j}^2(t)
	=
	\|\bm X_{\bm n,j}(t)-\bm U\|^2
	=
	a^2+R_{\rm e}^2-2aR_{\rm e}\,s_{\bm n,j}(t).
	\label{eq:distance}
\end{align}
Therefore, minimizing this distance is equivalent to maximizing
the cosine distance. For notation simplicity, we refer to this cosine distance as \emph{closeness}.

Let \(\phi_{\bm n,j}(t)\) denote the elevation angle above
the user's local horizontal plane, as shown in Fig. \ref{fig2}. Define its minimum elevation angle $0\leq\phi_{\min}<\frac{\pi}{2}$ used in practical deployments, where excessively large path loss is unacceptable.
A simple algebraic argument shows that its corresponding minimum allowable closeness equals
\begin{align}
c:=(R_e/a)\cos^2\phi_{\min}+
\sin\phi_{\min}
\sqrt{1-(R_e/a)^2\cos^2\phi_{\min}}.
\end{align}
Consequently, when $\phi_{\bm n,j}(t)\geq\phi_{\min}$, $c\leq s_{\bm n,j}(t)\leq 1.$
Under the assumptions mentioned above, $0<c\leq 1$. In particular, when $\phi_{\min}=0$, the threshold is $c=R_e/a$.

\subsection{From Orbital Phase to Projection Space}
The closeness metric $s_{\bm n,j}(t)$ in (\ref{eq:closeness}) can be further re-phrased as
\begin{align}
	s_{\bm n,j}(t)
	&=r_{\bm n}\cos(\theta_{\bm n,j}+\omega t)
	\nonumber\\
	&=r_{\bm n}\cos\theta_{\bm n,j}\cos(\omega t)-r_{\bm n}\sin\theta_{\bm n,j}\sin(\omega t), 
	\label{eq:closeness-expansion}
\end{align}
where $r_{\bm n}:=r{(\bm n})$. This expansion motivates the following state point representation of satellite locations.

\begin{definition}[State Point Representation]
	The corresponding static point of satellite $(\bm n,j)$ is defined as 
	\begin{equation}
		\bm p_{\bm n,j}
		:=(A_{\bm n,j}, B_{\bm n,j})^{\mathsf T}
		= r_{\bm n}(\cos\theta_{\bm n,j},-\sin\theta_{\bm n,j})^{\mathsf T}.
		\label{eq:point-map}
	\end{equation}
	Besides, the rotational query vector is defined as
	\begin{equation}
		\bm w(\varphi)
		=(\cos\varphi, \sin\varphi)^{\mathsf T},
		\qquad
		\varphi=\omega t.
		\label{eq:query-direction}
	\end{equation}
\end{definition}
Hence, any satellite, regardless of the orbit in which it is located, can be expressed in a new linear space with basis vector $\bm w(\varphi)$ or $\bm w(\omega t)$, as 
\begin{align}
s_{\bm n,j}(t)=\bm p_{\bm n,j}^{\mathrm T}\bm w(\omega t)=\bm p_{\bm n,j}^{\mathrm T}\bm w(\varphi).
\end{align}
Moreover, $\|\bm p_{\bm n,j}\|=r_{\bm n}\leq 1,$ where $r_{\bm n}$ depends on the orbit normal vector and user direction, rather than on satellite phase. 

\begin{figure}[!t]
	\centering
	\includegraphics[width=3.7in]{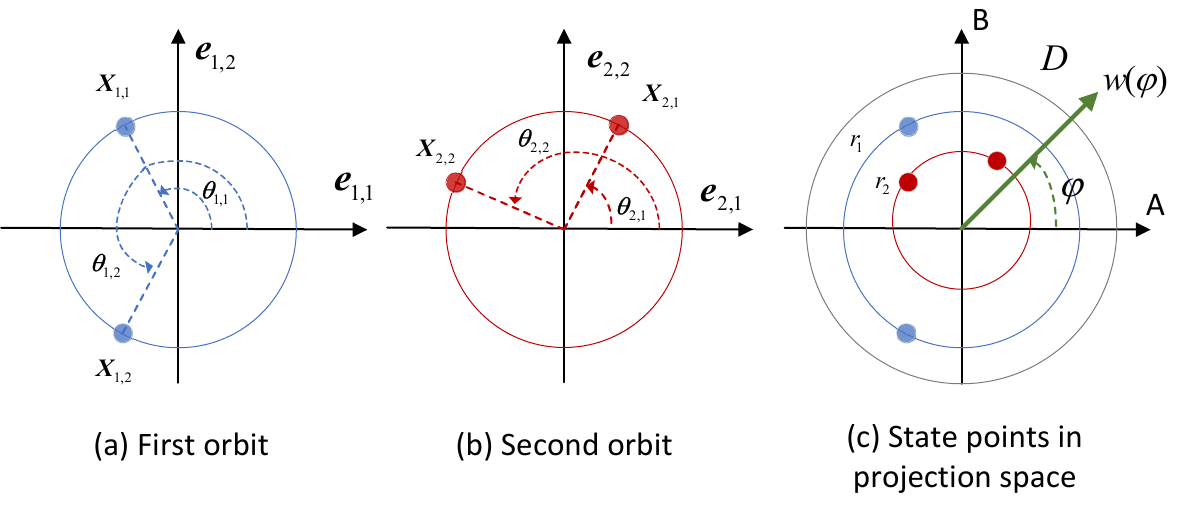}
	\caption{The left and middle sub-figures show the initial phases $\theta_{\bm n,j}$ for the two illustrated orbits, respectively. We first project a physical trajectory onto the user direction to obtain a sinusoidal functional. Then, store its cosine and sine coefficients as a static point in the projection space, as shown in the right sub-figure. The geometric quantity $r_j$ is the radius of the orbit circle inside the transformed disk of the projection space. The phase due to satellite motion increases by the same amount $\omega t$ on every orbit, which is realized by rotating the direction vector $\bm w(\omega t)$ or $\bm w(\phi)$.}
	\label{fig3}
\end{figure}

Thus, all satellites on one orbit are mapped to points on a circle of radius $r_{\bm n}$, while satellites on different orbits can be compared in the same projection space. Note that a point in this new projection space is a representation of satellite trajectories, rather than a projection of physical satellite positions onto a common geographic plane. The orbital phase and the orbit-dependent amplitude remain fixed in projection space over time when $\bm w(\omega t)$ is kept unchanged. Fig. \ref{fig3} shows an example on the transformation from physical space to projection one.

To sum up, there are two equivalent descriptions for satellite movement inside their orbits:
\begin{enumerate}
	\item In physical space, the satellites move clockwisely and the
	user is fixed.
	\item In projection space, all satellite points are fixed and the rotational query direction $\bm w(\omega t)$ rotates counterclockwise.
\end{enumerate}

Hence, we can utilize closeness metric in projection space to determine the best serving satellite for the typical user. In this paper, we consider the \emph{nearest-visible-satellite association} strategy. To be specific, 
let $\widetilde\Phi=\{\bm p_{\bm n,j}\}_{\bm n,j}$ denote the set of transformed satellite points in the projection space. 
The serving satellite is the one has the maximum closeness, namely, 
\begin{align}
	M(t)=\sup_{\bm n, j} s_{\bm n, j}(t)=\max_{\bm p\in\widetilde\Phi}
	\bm p^{\mathsf T}\bm w(\omega t),
\end{align}
with \(M(t)=-\infty\) if there are no satellites. Besides, the user is connected at time $t$\ if $M(t)>c$.

\subsection{The Induced Cox Process in Projection Space}
Recall that conditional on the parent orbits, satellite phases on different orbits are independent
Poisson processes, each with intensity $\mu\,\dd\theta/(2\pi)$. 
At any fixed time $t$, adding common $-\omega t$  to every phase point leaves the uniform phase intensity unchanged. 
Since the position transformation in (\ref{eq:point-map}) changes only the radius $r$ of the satellite points and not the initial phase, the image of the phase process in projection space is still a Poisson point process on the circle of radius $r$ with uniform angular distribution and the same intensity $\mu$. To simplify exposition, the transformed points in projection space is named \emph{state point process}, since each point represents a satellite with specified parameters. The point corresponding to phase $\theta$ on an orbit with transformed radius $r$ is rewritten as 
\begin{equation}
	\bm p_r(\theta)=
	r(\cos\theta,-\sin\theta)^{\mathsf T}.
	\label{eq:phase-to-disk}
\end{equation}
Furthermore, we introduce the new variable $\alpha=-\theta\bmod 2\pi$. The mapping $\theta\mapsto\alpha$ preserves the uniform angular measure, so the image of the phase process is a Poisson point process with a uniform angular distribution and the same intensity in the $\alpha$-coordinate.

Recall that for an orbit with normal vector $\bm n$, its transformed radius is $r=\sqrt{1-(\bm u^{\mathsf T}\bm n)^2}$.
Accordingly, we have the following statistical distribution for state point process.
\begin{proposition}[Isotropic radius law]
	\label{prop:radius}
	The transformed radius has probability density
	\begin{equation}
		f_r(\rho)=\frac{\rho}{\sqrt{1-\rho^2}},
		\qquad 0\leq \rho\leq 1.
		\label{eq:radius-density}
	\end{equation}
\end{proposition}

\begin{proof}
	For a uniform normal vector $\bm n$ on $\sphere$, $\bm u^{\mathsf T}\bm n$ is uniformly distributed in $[-1,1]$. Hence, $Y=|\bm u^{\mathsf T}\bm n|$ is uniform on $[0,1]$. Since $r=\sqrt{1-Y^2}$,
	\begin{align}
		\Prob\{r\leq \rho\}=
		\Prob\{Y\geq\sqrt{1-\rho^2}\}
		=1-\sqrt{1-\rho^2}.
	\end{align}
	Differentiation of this probability gives \eqref{eq:radius-density}.
\end{proof}

\begin{proposition}[State point process on one orbit]
	\label{prop:ho-one-orbit-ppp}
	Conditional on a parent orbit having radius $r$ in projection space, $\widetilde{\Phi}_r$ is a Poisson point process on this orbit with intensity measure
	\begin{equation}
		\widetilde\Lambda_r(D)=
		\frac{\mu}{2\pi}
		\int_0^{2\pi}
		\mathbf 1\!\left\{
		(r\cos\alpha,r\sin\alpha)\in D
		\right\}\,\mathrm d\alpha,
		\ D\in\mathcal B(\mathbb R^2).
		\label{eq:ho-orbit-intensity}
	\end{equation}
\end{proposition}

\begin{proof}
	Let $\bm p_r(\alpha)=r(\cos\alpha,\sin\alpha)^{\mathsf T}$ be the points transformed from satellite phase process with intensity $\mu\,\dd\alpha/(2\pi)$. For any Borel set $D$, the intensity measure is expressed as
	\begin{align}
\frac{\mu}{2\pi}\int_0^{2\pi}\mathbf 1\{\bm p_r(\alpha)\in D\}\,\mathrm d\alpha,
	\end{align}
	which is exactly \eqref{eq:ho-orbit-intensity}.
\end{proof}

\begin{proposition}[Aggregate state point process]
	\label{thm:ho-aggregate-cox}
	Conditional on the parent orbit plane process $\Xi$, the aggregate state point process $\widetilde\Phi=\sum_{{\bm n}\in \Xi}\widetilde\Phi_{r_{\bm n}}$  is a Poisson point process with random intensity measure
	\begin{equation}
		\widetilde\Lambda_{\Xi}(D)=\sum_{{\bm n}\in \Xi}\widetilde\Lambda_{r_{\bm n}}(D),
		\quad D\in\mathcal B(\mathbb R^2),
		\label{eq:ho-random-intensity}
	\end{equation}
	 Since $\widetilde\Lambda_{\Xi}$ is a random measure driven by $\Xi$, $\widetilde\Phi$ is a Cox point process with intensity measure $\widetilde\Lambda(D)=\E[\widetilde\Lambda_{\Xi}(D)]$.
\end{proposition}

\begin{proof}
	Conditional on the parent plane orbit process, the
	satellite phase processes on distinct parent orbits
	are independent. By Proposition~\ref{prop:ho-one-orbit-ppp}, their images in projection space are therefore independent Poisson point processes with intensity measures $\{\widetilde\Lambda_{r_{\bm n}}\}_{\bm n}$. The superposition of independent Poisson point processes is still Poisson, with intensity measures equal to the sum of their intensity ones.
	Hence, conditional on the parent plane orbit process $\Xi$, $\widetilde\Phi$ is Poisson with intensity $\widetilde\Lambda_{\Xi}(D)$ indicated by \eqref{eq:ho-random-intensity}. Since this
	intensity measure is random, $\widetilde\Phi$ is a Cox
	process by the definition.
\end{proof}

\subsection{Void Event and Probability}
To evaluate the handover procedure, it is essential to determine the handover candidate satellites under the condition that there are no other competitors. To this end, we first define disk cap as below.
\begin{definition}[Disk cap]
	For a rotational query vector $\bm w$ and closeness threshold $z$, define the open disk cap
	\begin{equation}
		\capset(\bm w,z)
		=
		\{\bm p\in\disk:\bm p^{\mathsf T}\bm w>z\},
		\label{eq:cap-definition}
	\end{equation}
	where $\disk$ denotes the domain in projection space in which all transformed points lie, and its radius is exactly $1$.
\end{definition}

\begin{figure}[!t]
	\centering
	\includegraphics[width=2.5in]{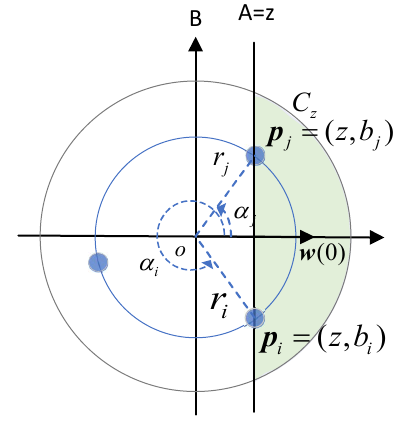}
	\caption{Local geometry of a connected handover.
		The switching satellites share the supporting
		line $A=z$, with $z>c$. The shaded cap in green must contain
		no other satellites except two candidate satellites. 
		The inner circle in blue represents a orbits in the projection space. The outer circle in grey represents the boundary of the projection space. 
	}
	\label{fig4}
\end{figure}
A cap is the portion of the unit disk $\disk$ having larger closeness than a threshold when rotation direction is given. The line $\bm p^{\mathsf T}\bm w=z$ is its boundary chord, and $\bm w$ points into the cap, as shown in Fig. \ref{fig4}. 
The closeness $\bm p^{\mathsf T}\bm w$ is the projection of point $\bm p$ onto
the rotational direction vector $\bm w$. Therefore,  the cap consists of points whose closeness is larger than $z$. 
Additionally, for a fixed visibility threshold $c$, the user is covered at time $t$ if and only if,
\begin{equation}
	\widetilde\Phi\bigl(\capset(\bm w(\omega t),c)\bigr)> 0,
	\label{eq: cov-cap}
\end{equation}
namely, it has at least one satellite with closeness exceeding $c$ exactly.
\begin{proposition}[Cox void functional]
	\label{prop:void}
	For every Borel set $D\subseteq\disk$,
	\begin{align}
		\voidfun(D)
		&:=
		\Prob\{\widetilde\Phi(D)=0\}\notag\\
		&=
		\exp\left[
		-\lambda\int_0^1
		\left(1-e^{-\widetilde\Lambda_r(D)}\right)
		f_r(\rho)\,\dd \rho
		\right],
		\label{eq:cox-void}
	\end{align}
	where $\lambda$ is the intensity of orbit plane process.
\end{proposition}

\begin{proof}
	Conditional on $\Xi$, the void probability is
	\begin{align}
\Prob\{\widetilde\Phi(D)=0\mid\Xi\}
=\exp\!\left[
-\sum_{\bm n\in\Xi}\widetilde\Lambda_{r_{\bm n}}(D)
\right].
\end{align}

Taking expectation over the Poisson parent process $\Xi$ and using its probability generating functional gives

\begin{align}
	\Prob\{\widetilde\Phi(D)=0\}
	&=
	\exp\!\left[
	-\frac{\lambda}{4\pi}
	\int_{\mathbb S^2}
	\left(
	1-e^{-\widetilde{\Lambda} _{r_{\bm n}}(D)} 
	\right)
	\,\dd\sigma(\bm n)
	\right]\notag\\
	&=
	\exp\!\left[
	-\lambda\,
	\mathbb E_r\!\left(
	1-e^{-\widetilde{\Lambda}_r(D)}
	\right)
	\right].
\end{align}

	The last equation is because the normalized spherical integral is the expectation of $r$. Then, writing the expectation with respect to the density $f_r(\rho)$ proves \eqref{eq:cox-void}.
\end{proof}

For a special rotational query direction $\bm w(0)=(1,0)^{\mathsf T}$, define cap
\begin{equation}
	C_z:=\capset(\bm w(0),z),
	\qquad 0<z<1.
	\label{eq:cap-Cz}
\end{equation}
Then, we have the following corollary. 
\begin{corollary}
The void probability of the cap $C_z$ is
\begin{align}
\voidfun(C_z)=
	\exp\left[
	-\lambda\int_z^1
	\left(
	1-\exp\left\{-\frac{\mu}{\pi}\arccos\frac{z}{\rho}\right\}
	\right)
	f_r(\rho)\,\dd \rho
	\right].
	\label{eq:Vz-explicit}
\end{align}
\end{corollary}

\begin{proof}
Suppose a point on the circle of radius $r$ in projection space has
the form $\bm p=r(\cos\alpha,\sin\alpha)^{\mathsf T}.$
It belongs to $C_z$ only when $\bm w(0)^{\mathsf T}\bm p=r\cos\alpha>z.$
There are no such points if $r\leq z$. If $r>z$, the condition selects the arc range $-\arccos(z/r)<\alpha<\arccos(z/r),$. Therefore, according to \eqref{eq:ho-orbit-intensity},
\begin{equation}
	\widetilde{\Lambda}_r(C_z)
	=
	\begin{cases}
		\displaystyle \frac{\mu}{\pi}\arccos\frac{z}{r},
		& r>z,\\[6pt]
		0,& r\leq z.
	\end{cases}
	\label{eq:single-cap-fraction}
\end{equation}

Bring this intensity measure into \eqref{eq:cox-void} to obtain \eqref{eq:Vz-explicit}.
\end{proof}

In this paper, we define the coverage probability as the probability that at least one satellite is visible. With Cox void probability derived above, we express coverage probability as
\begin{equation}
	p_{\mathrm{cov}}=1-\voidfun(C_c).
	\label{eq:coverage}
\end{equation}
Note that this coverage is valid snapshot performance for any time, because the state point process is time stationary.

\subsection{Palm Void Probabilities with Tagged Orbits} 
When a handover takes place in a given Borel set $D\subseteq\disk$ among satellites locating on one or more parent orbits, the left satellites shall remain in the same configuration outside set $D$. It is convenient to take the Palm distribution and its reduced version to study the void probability.

\begin{proposition}[Reduced-Palm void probability for tagged orbits]
	\label{prop:tagged}  
	Suppose one or more satellites are selected for handover from a parent orbit
	with transformed radius $r$, and remove the selected satellites.
	Under the corresponding reduced Palm distribution, the
	probability that all remaining satellites avoid \(D\) is
	\begin{equation}
		Q_1(D;r)
		=
		\voidfun(D)e^{-\widetilde{\Lambda}_r(D)}.
		\label{eq:Qone}
	\end{equation}
	If selected satellites belong to two distinct parent orbits with
	transformed radii \(r_1\) and \(r_2\), then
	\begin{equation}
	Q_2(D;r_1,r_2)
		=
		\voidfun(D)
		\exp\!\left[
		-\widetilde{\Lambda}_{r_1}(D)-\widetilde{\Lambda}_{r_2}(D)
		\right].
		\label{eq:Qtwo}
	\end{equation}
\end{proposition}

\begin{proof}
	The reduced Palm law of a Poisson process leaves the unselected process unchanged in distribution. Applied to the parent process, the untagged orbits
	remain an independent copy of the original parent process. They produce the factor $\voidfun(D)$.
	
	Applied to the satellite phase process on a tagged
	orbit, removing the selected satellite or satellites
	leaves an independent Poisson phase process with the
	original mean measure. Its void probability in $D$
	is $e^{-\widetilde{\Lambda}_r(D)}$.
	
	Two distinct tagged orbits contribute independent
	such factors. Multiplication all of them proves \eqref{eq:Qtwo}.
	
\end{proof}
Note that if two selected satellites belong to the same parent orbit, the
tagged orbit factor $e^{-\widetilde{\Lambda}_r(D)}$ appears only once.

\section{Connected Handover Intensity}
\label{subsec:handover-definition}
In this section, we discuss connected handover and show how to obtain its intensity. The term \emph{connected handover}, or \emph{handover} for short, occurs when the serving satellite changes while the common closeness is strictly above c. 
Specifically, a link is switched from source satellite $i$ to target satellite $j$ at time $t_0$ if the satellite $i$ is the unique serving node immediately before $t_0$, and satellite $j$ is the unique serving node immediately after $t_0$. 
A transition from the disconnected state to a connected state is a connection recovery instead of a handover, which is out of our concern.

We work under the non-degeneracy assumptions on point processes in
projection space: the process has almost surely finitely many points, distinct satellites have distinct state points almost surely. 
Moreover, we assume that triple satellites have the same closeness with probability zero. 
In the orbit model specified below, these assumptions are from the continuous phase distributions and the diffuse radius law of Cox point process.

\subsection{Local Handover Geometry} 
Since the satellite phase processes are invariant under a common phase
shift, the handover model is stationary in time. 
Without loss of generality, we may therefore study the handover procedure at the reference time $t=0$ such that $\bm w(0)=(1,0)^{\mathsf T}$, as shown in Fig. \ref{fig4}.  For a candidate satellite of radius $r$ and phase $\alpha$, its closeness at $t=0$ is $z=r\cos\alpha$, and its second coordinate is $b=r\sin\alpha$. For $c<z$, define $q(r,z)=\sqrt{r^2-z^2}.$ Thus, 
\begin{align}
	\left|\frac{\mathrm dz}{\mathrm d\alpha}\right|
	=r|\sin\alpha|
	=q(r,z).
	\label{dz/da}
\end{align}
\begin{figure}[!t]
	\centering
	\includegraphics[width=3.6in]{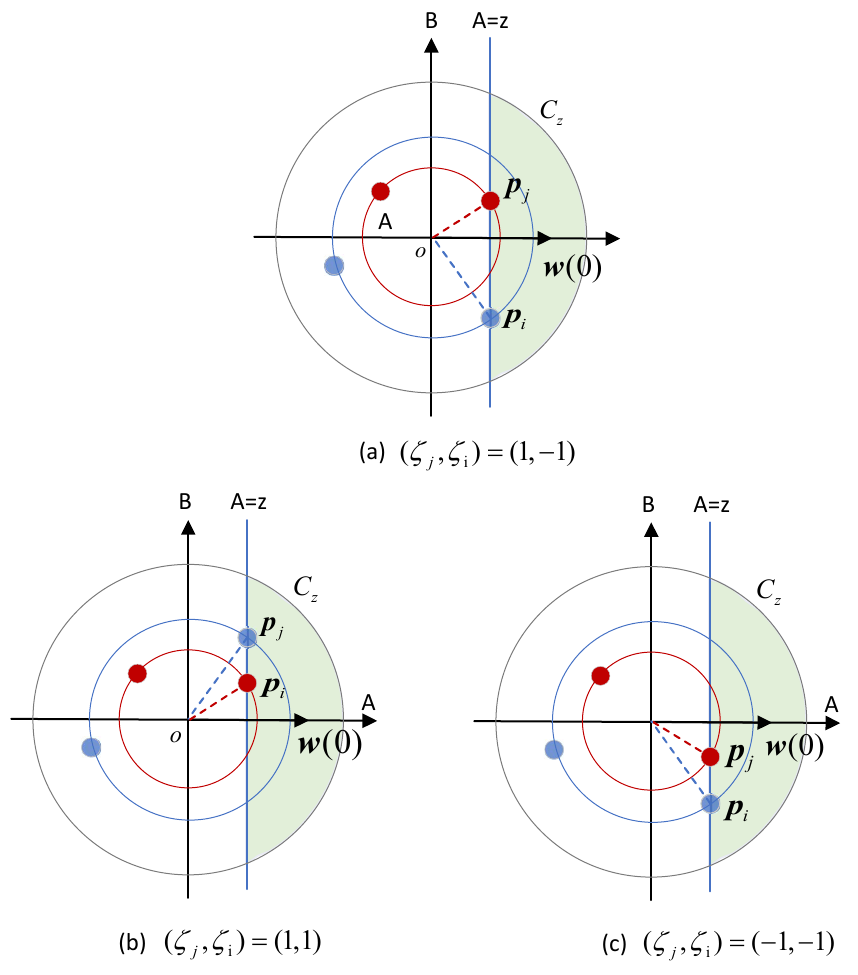}
	\caption{Three candidate combinations of satellite positions in the projection space, where two satellites are located in different orbits: (a)$(\zeta_j,\zeta_i)=(1,-1)$;(b) $(\zeta_j,\zeta_i)=(1,1)$;(c) $(\zeta_j,\zeta_i)=(-1,-1)$.}
	\label{fig5}
\end{figure}
The handover happens when both outgoing and incoming satellites have the same closeness, say $z$. 
In the following, the subscript $\bm n$ is dropped for brevity when no confusion arises.
Denote these satellites at this time $t=0$ as $\bm p_i=(z,b_i)$, and $\bm p_j=(z,b_j)$ with radii $r_i$ and $r_j$, respectively. When two satellites are on the same orbit, $r_i=r_j$, whereas $r_i\neq r_j$ for satellites on different orbits. Their phases are expressed as $\alpha(r_j,z)=\arccos(z/r_j),\alpha(r_i,z)=2\pi-\arccos(z/r_i),$ respectively.  Hence, the second coordinates of the two points are respectively expressed as
\begin{align}
	b_{j}=\zeta_j q(r_j,z), 
	\quad b_{i}=\zeta_i q(r_i,z),
	\quad \zeta_i,\zeta_j\in\{-1,+1\},
	\label{eq:ho-phase-branches}
\end{align} 
where auxiliary variables $\zeta_i,\zeta_j$ are used to indicate two candidate satellite locations, as shown in Fig. \ref{fig5}. Different $\zeta_i,\zeta_j$ combinations correspond to different candidate locations in the projection space. The incoming satellite $j$ is assumed to have a larger second coordinate than the outgoing satellite $i$.
With $\bm p_i,\bm p_j$ fixed, the direction vector rotated through $\varphi=\omega t$,  their closeness difference is computed as
\begin{align}
		g_{ij}(t)&:=s_j(t)-s_i(t)=
	\bm p_j^{\mathsf T}\bm w(\omega t)
	-
	\bm p_i^{\mathsf T}\bm w(\omega t)\notag\\
	&=
	(b_j-b_i)\sin(\omega t),
	\label{eq:ho-relative-score}
\end{align} 
and $g_{ij}(0)=0$. The differential of $g_{ij}(t)$ at time $t=0$ is derived as 
\begin{align} 
{\dot g}_{ij}(0):=\left.\frac{\mathrm d}{\mathrm dt}
\bigl(s_j(t)-s_i(t)\bigr)\right|_{t=0}
=\omega(b_j-b_i)\notag\\
=\omega\bigl(\zeta_j q(r_j,z)-\zeta_i q(r_i,z))\bigr).
\end{align}
When the handover is from the outgoing satellite at $\bm p_i$  to incoming one at $\bm p_j$, we have $b_j>b_i$, i.e., ${\dot g}_{ij}(0)>0$. In this handover direction, the incoming satellite has a larger closeness than the outgoing one, making it a promising handover target.  Thus, the positive relative closeness speed is determined by 
\begin{align}
\omega\bigl[\zeta_j q(r_j,z)-\zeta_i q(r_i,z)]_+,
	\label{eq:g_positive}
\end{align}
where \([x]_+=\max\{x,0\}\). Equivalently, a handover takes place only if  (\ref{eq:g_positive}) is positive.
 In Fig. \ref{fig4} and Fig. \ref{fig5}, the equal-closeness line is $\ell_z=\{(A,B):A=z\}$. The handover is equivalent to the event that no other satellite lies in $C_z$, conditional on the two tagged switching satellites lying on $\ell_z$. 
 Fig. \ref{fig4} shows the satellite positions when both satellites are located on the same orbit, while Fig. \ref{fig5} shows three candidate position combinations for the case where the two satellites are located on different orbits.

To summarize the above discussion, a connected handover from outgoing $\bm p_i$ to incoming $\bm p_j$ at a time $t$ almost surely, when following three conditions are satisfied
\begin{align}
	 &s_i(t)=s_j(t)=z>c,
	\label{eq:ho-tie}\\
	 &s_{ k}(t)<z,
	\quad\forall { k}\notin\{i,j \},
	\label{eq:ho-dominance}\\
	 &\dot g_{ij}(t)>0.
	\label{eq:ho-orientation}
\end{align}
The first condition maintains connectivity of two switching satellites at the handover event. The second makes the switching pair jointly optimal. The third
specifies the direction of the change: the incoming satellite
overtakes the outgoing satellite.
Connection establishment and connection loss at the threshold $c$ are not regarded as connected handovers. 

\subsection{Handover Point Process and Intensity}
According to the handover conditions discussed above, we define the handover point process by
\begin{align}
	&\mathcal H:=\sum_{i\ne j} \sum_{\substack{t\in\mathbb R:\,g_{ij}(t)=0}}
	\delta_t\notag\\
	&\times\mathbf 1\!\left\{
	s_i(t)=s_j(t)=z>c,s_k(t)<z, \forall k\notin\{i,j\},	\dot g_{ij}(t)>0
	\right\}.
	\label{eq:ho-point-process}
\end{align}
where \(\delta_t\) is Dirac measure concentrated at time $t$. The sum is over any ordered pairs: $\bm p_i$ is outgoing satellite and $\bm p_j$ incoming one. At any handover, reversing the
pair reverses the sign of \(\dot g_{ij}\), so exactly
one ordering contributes when two satellites switch.

The handover point process $\mathcal H$ is stationary, because a time shift rotates all state points by a common angle. Therefore, the state point process is rotation invariant, and the handover rule is preserved by this time shift.
Consequently, its intensity, when finite, satisfies
\begin{equation}
	\nu_{\mathrm{HO}}
	=
	\frac{\mathbb E[\mathcal H((0,T])]}{T},
	\qquad T>0.
\end{equation}

\begin{theorem}[Connected handover intensity]
	\label{thm:ho-rice}
	Let \(g_{ij}(t)=s_j(t)-s_i(t)\).
	Under the above nondegeneracy assumptions and the
	regularity conditions needed for the generalized
	expectation below, the connected handover intensity is given by
	\begin{align}
		\nu_{\mathrm{HO}}
		=
		\mathbb E\Bigg[
		\sum_{i\ne j}
		&\delta_0\!\bigl(g_{ij}(0)\bigr)
		[\dot g_{ij}(0)]_+
		\notag\\[-2pt]
		&\times
		\mathbf 1\!\left\{
		s_i(0)>c,\ 
		s_k(0)<s_i(0),\
		\forall k\notin\{i,j\}
		\right\}
		\Bigg],
		\label{eq:ho-rice}
	\end{align}
where \([x]_+=\max\{x,0\}\), and \(\delta_0\) denotes the Dirac delta measure at zero.

\end{theorem}

\begin{proof}
At every handover time $t_0$, we have $	g_{ij}(t_0)=0, \dot g_{ij}(t_0)\ne 0$ for  a realization and an ordered pair $(i,j)$. We first explain how positive switching are counted before considering left handover conditions. The number of positive switching on $(0,T]$ is
\begin{align}
N_{ij}(0,T]:=\sum_{\substack{t_0\in(0,T]:\\
	g_{ij}(t_0)=0}}
\mathbf 1\{\dot g_{ij}(t_0)>0\}.
\end{align}
	
Since positive switching can be equivalently expressed as $\mathbf 1\{\dot g_{ij}(t_0)>0\}=\frac{[\dot g_{ij}(t_0)]_+}{|\dot g_{ij}(t_0)|},$ and hence
\begin{equation}
		N_{ij}(0,T]
		=
		\sum_{\substack{t_0\in(0,T]:\\
				g_{ij}(t_0)=0}}
		\frac{[\dot g_{ij}(t_0)]_+}
		{|\dot g_{ij}(t_0)|}.
		\label{eq:positive-crossing-sum}
	\end{equation}
	
	Denote $\delta_{t_0}(\mathrm dt)$  as the unit Dirac
	measure at the time \(t_0\). It has the integral property $\int_{(0,T]} h(t)\,\delta_{t_0}(\mathrm dt)=h(t_0),\ t_0\in(0,T]$, which make the $N_{ij}(0,T]$ become
	\begin{align}
		N_{ij}(0,T]
		&=
		\sum_{\substack{t_0\in(0,T]:\\
				g_{ij}(t_0)=0}}
		\int_{(0,T]}
		\frac{[\dot g_{ij}(t)]_+}
		{|\dot g_{ij}(t_0)|}
		\,\delta_{t_0}(\mathrm dt)
		\notag\\
		&=
		\int_{(0,T]}
		[\dot g_{ij}(t)]_+
		\left(
		\sum_{\substack{t_0\in(0,T]:\\
				g_{ij}(t_0)=0}}
		\frac{\delta_{t_0}(\mathrm dt)}
		{|\dot g_{ij}(t_0)|}
		\right).
		\label{eq:crossing-dirac-time}
	\end{align}
	
Using a simple change of variables identity for the delta function, we obtain
	\begin{equation}
		\delta_0(g_{ij}(t))\,\mathrm dt
		=
		\sum_{t_0:\,g_{ij}(t_0)=0}
		\frac{\delta_{t_0}(\mathrm dt)}
		{|\dot g_{ij}(t_0)|},
		\label{eq:delta-time-change}
	\end{equation}
	in the sense of distribution.
Substitute (\ref{eq:delta-time-change}) into (\ref{eq:crossing-dirac-time}), we obtain
\begin{align}
N_{ij}(0,T]
=
\int_{(0,T]}
\delta_0(g_{ij}(t))
[\dot g_{ij}(t)]_+\,\mathrm dt.
\end{align}

Define $F(t)=\mathbf 1\!\left\{ s_i(t)>c,\ s_k(t)<s_i(t),\ \forall k\notin\{i,j\}\right\}$
as the dominance and coverage conditional indication function.
With the same argument, with an additional non-negative
measurable function $F(t)$, we have the relation
	\begin{equation}
		\sum_{\substack{t_0\in(0,T]:\\
				g_{ij}(t_0)=0,\ 
				\dot g_{ij}(t_0)>0}}
		F(t_0)
		=
		\int_{(0,T]}
		\delta_0(g_{ij}(t))
		[\dot g_{ij}(t)]_+F(t)\,\mathrm dt.
		\label{eq:weighted-positive-crossings}
	\end{equation}
Then, summing over ordered pairs counts each connected handover exactly once:
	\begin{equation}
		\mathcal H((0,T])
		=
		\sum_{i\ne j}
		\int_{(0,T]}
		\delta_0(g_{ij}(t))
		[\dot g_{ij}(t)]_+
		F(t)\,\mathrm dt.
		\label{eq:handover-pathwise-count}
	\end{equation}
	
Taking expectations by Tonelli's theorem, we arrive at 
\begin{align}
\mathbb E[\mathcal H((0,T])]=\int_0^T \mathbb E\!\left[
\sum_{i\ne j}
\delta_0(g_{ij}(t))
[\dot g_{ij}(t)]_+F(t)
\right]\mathrm dt.
\end{align}
With the stationarity property of handover point process,
\begin{align}
\mathbb E[\mathcal H((0,T])]=T\mathbb E\!\left[
\sum_{i\ne j}
\delta_0(g_{ij}(0))
[\dot g_{ij}(0)]_+F(t)
\right].
\end{align} 
Dividing by \(T\) proves this theorem.
\end{proof}

Theorem \ref{thm:ho-rice} implies a way to compute the handover intensity. Conditional on dominance and coverage, we simply count the number of candidate pairs when a positive switching takes place at a reference time, which requires computing the conventional factorial intensity. In what follows, we compute the handover intensities for the same-orbit and different-orbit cases, respectively.
\subsection{Same-Orbit Handover}
Suppose general incoming and outgoing satellites have transformed radii $r_i, r_j$, respectively. 
Satellite phases on the parent orbit in the projection space form a Poisson point process with intensity $\frac{\mu}{2\pi}\,\mathrm d\alpha$. 
According to (\ref{dz/da}), the intensity expressed in terms of the coordinate $z$ becomes $\frac{\mu}{2\pi q(r,z)}\,\mathrm dz.$ 
Thanks to the Poisson property, all ordered pair of distinct point has  factorial intensity 
\begin{align}
\frac{\mu^2}{4\pi^2q_iq_j}\,
\mathrm dz_i\,\mathrm dz_j
\end{align}
where $q_i:=q(r_i,z_i)=\sqrt{r_i^2-z_i^2}$ and $q_j:=q(r_j,z_j)=\sqrt{r_j^2-z_j^2}$.
To compute explicitly handover intensity, we directly apply the expression in (\ref{eq:ho-rice}) at the reference time $t=0$. The equal closeness condition is imposed by $\delta_0(g_{ij}(0))=\delta_0(z_j-z_i)$, and $[\dot g_{ij}(0)]_+=\omega(\zeta_j q(r_j,z_j)-\zeta_i q(r_i,z_i))=2\omega(q(r_j,z_j)+q(r_i,z_i))$, where $(\zeta_j,\zeta_i)=(1,-1)$ due to symmetric locations with respect to first coordinate in the projection space. 
Thus, integrating out one of the two closeness variables leads to $z_i=z_j=z$, $q_i=q_j=q$. 
Then, the positive switching rate contribution per unit time for a pair of satellites become
\begin{equation}
	\frac{\mu^2}{4\pi^2q^2}\,
	\mathrm dz*2\omega q=
	\frac{\omega\mu^2}{2\pi^2q}
	\,\mathrm dz.
	\label{eq:ho-branch-crossing-rate}
\end{equation}

Then, we obtain the handover intensity by the following theorem for the same-orbit handover.
\begin{theorem}[Same-orbit handover intensity]
The same-orbit connected handover intensity is given by
	\begin{equation}
		\nu^{\mathrm s}_{\mathrm{HO}}
		=
		\frac{\omega\lambda \mu^2}{2\pi^2}
		\int_c^1\int_z^1
		\frac{f_r(\rho)Q_{1}(C_z;\rho)}
		{\sqrt{\rho^2-z^2}}
		\,\mathrm d\rho\,\mathrm dz,
		\label{eq:ho-same-intensity}
	\end{equation}
\end{theorem}

\begin{proof}
The selected parent orbit intensity is $\lambda f_r(\rho)\,\mathrm d\rho$ in projection space. For two satellites on the same parent orbit, the factorial intensity of an ordered pair of distinct satellites is given by \eqref{eq:ho-branch-crossing-rate}. By Proposition~\ref{prop:tagged}, the probability that no remaining satellite lies in $C_z$ is $Q_{1}(C_z;\rho)$. 
This includes the remaining satellites on the selected parent orbit. Combining $q=\sqrt{\rho^2-z^2}$ with the ordered pair satellite intensity, the parent orbit intensity, and the tagged orbit void probability, and then integrating over $c<z<\rho<1$,  yields \eqref{eq:ho-same-intensity}.
\end{proof}
\subsection{Different-Orbit Handover}
Following the same argument as in the same-orbit case, for the outgoing and incoming satellites on two distinct parent orbits with transformed radii $r_i$ and $r_j$, respectively, define $q_i=\sqrt{r_i^2-z^2}$ and $q_j=\sqrt{r_j^2-z^2}$ at their common closeness $z$. Unlike the same-orbit case, $\zeta_i$ and $\zeta_j$ need not agree. 
	
\begin{theorem}[Different-orbit handover intensity] The different-orbit connected handover intensity is 
	\begin{align}
		\nu^{\mathrm d}_{\mathrm{HO}}&=  
		\frac{\omega\lambda^2\mu^2}{4\pi^2}
		\int_c^1\int_z^1\int_z^1
		\frac{f_r(\rho_i)f_r(\rho_j)}
		{q_iq_j}
		\notag\\[-1mm]
		&\times
		\bigl(2\max\{q_i,q_j\}\bigr)
		Q_2(C_z;\rho_i,\rho_j)
		\,\mathrm d\rho_j\,\mathrm d\rho_i\,\mathrm dz,
		\label{eq:ho-different-intensity}
	\end{align}
\end{theorem}

\begin{proof}
	The factorial intensity of an ordered pair of
	distinct parent orbits with respective radii $\mathrm dr_i$ and $\mathrm dr_j$ in projection space is $	\lambda^2 f_r(\rho_i)f_r(\rho_j)\, \mathrm d\rho_i\,\mathrm d\rho_j.$

As in the same-orbit case, selecting one satellite on each parent orbit yields the intensity for equal closeness $\frac{\mu^2}{4\pi^2q_iq_j}\,\mathrm dz.$
Take the switching direction  into account,  this intensity is multiplied by the positive factor $\omega\bigl[\zeta_j q_j- \zeta_i q_i\bigr]_+$, which consists of four terms:
\begin{align}
[q_j-q_i]_+,\quad 0,\quad q_i+q_j,\quad [q_i-q_j]_+.
\end{align}
Their sum is $q_i+q_j+|q_j-q_i|=2\max\{q_i,q_j\}$.

The reduced Palm probability that no remaining satellite dominates the pair satellites is $Q_2(C_z,\rho_i,\rho_j)$. Multiplication with ordered pair terms, then integrating over $\rho_i,\rho_j>z>c$ yields
	\eqref{eq:ho-different-intensity}.
\end{proof}

The same-orbit and different-orbit cases are disjoint and exhaust all ordered pairs of distinct satellites. Hence, the previously defined handover point process intensity is
\begin{align}
v_{\mathrm{HO}}=v_{\mathrm{HO}}^{\mathrm{s}}
+v_{\mathrm{HO}}^{\mathrm{d}}.
\end{align} 
This intensity is also the unconditional per-unit-time rate of serving-satellite changes whose common closeness exceeds $c$, ignoring connection recovery and loss.

\section{Numerical Validation}
\subsection{A Toy Example: Two Satellites on One Orbit}
\label{subsec:toy-handover-check}
As a check of handover intensity, we consider a single orbit of fixed radius $r$ carrying exactly two satellites. Their phases are independent and
uniform on $[0,2\pi)$. Assume $0<c<r$. Let $\Delta\in[0,\pi]$ be the smaller angular separation between the two satellites. It is uniform
on $[0,\pi]$. 
As the rotational query vector $\bm w$ completes one full rotation, the closeness values of the two satellites become equal twice: once on the outward bisector of the two satellite directions, where their common closeness is $z_*=r\cos(\Delta/2),$
and once on the opposite bisector, where their common closeness is $-z_*$. 

Since $c>0$, only the outward bisector switching is under coverage. It is a handover only when $z_*>c$, or equivalently $\Delta<2\arccos(c/r).$
The rotation rate is $\omega/2\pi$. Since $\Delta \sim \mathrm{Unif}[0,\pi]$, we can count the handover rate as
\begin{align}
	\nu_{\mathrm{toy}}
	&=
	\frac{\omega}{2\pi}
	\mathbb{P}\!\left\{
	\Delta<2\arccos(c/r)
	\right\}
	\nonumber\\
	&=
	\frac{\omega}{2\pi}
	\frac{2\arccos(c/r)}{\pi}
	=
	\frac{\omega}{\pi^2}\arccos(c/r).
	\label{eq:toy-direct-rate}
\end{align}

We now compute the rate from the handover intensity formulation. At a positive switching, the outgoing and incoming satellites have $$\omega[b_j-b_i]_+=2\omega q(r,z), \ q(r,z)=\sqrt{r^2-z^2}.$$
There are no other satellites, so the dominance factor is one. The factorial count of ordered pair of different satellite is 2, and each satellite has the intensity expressed in closeness $z$ is $1/(2\pi q(r,z))$.
Hence, the handover intensity is given by
\begin{align}
	\nu_{\mathrm{toy}}
	&=
	2\int_c^r
	\left(\frac{1}{2\pi q(r,z)}\right)^2
	2\omega q(r,z)\,\dd z
	\nonumber\\
	&=
	\frac{\omega}{\pi^2}
	\int_c^r \frac{\dd z}{\sqrt{r^2-z^2}}
	=
	\frac{\omega}{\pi^2}\arccos(c/r),
	\label{eq:toy-flux-rate}
\end{align}
which agrees with \eqref{eq:toy-direct-rate}.

\subsection{Numerical Simulation}
We evaluate the handover intensity from the viewpoint of a fixed
typical ground user using a 3-D Earth-satellite geometry. The Earth is modeled as a sphere of radius $R_{\rm e}=6371~\mathrm{km}$. The satellite constellation is generated as a Cox point process.
The number of parent orbital planes $N_{\mathrm{orb}}\sim\operatorname{Poisson}(\lambda)$. 
Each orbital plane normal vector is sampled isotropically on the unit sphere.
Conditioned on a parent orbit, its number of satellites is $N_{\mathrm{sat}}\sim\operatorname{Poisson}(\mu),$
and the satellite initial phases are independent and uniformly
distributed on $[0,2\pi)$. All satellites have the same orbital radius
and angular speed.

The serving satellite is the visible one with the smallest user-satellite range, where visibility requires an elevation angle of at least $\phi_{\min}=10^\circ$. A handover is recorded when the serving satellite changes. Each switching is classified as a same-orbit handover if the two switched serving satellites belong to the same parent orbit, and as a different-orbit handover otherwise. 
The simulated intensity is the mean event count per realization divided by
the orbital period. 
The Monte Carlo simulation estimates are averaged over $N_{\mathrm{MC}}=3000$ independent Cox-process realizations.
\begin{figure}[!t]
	\centering
	\includegraphics[width=2.5in]{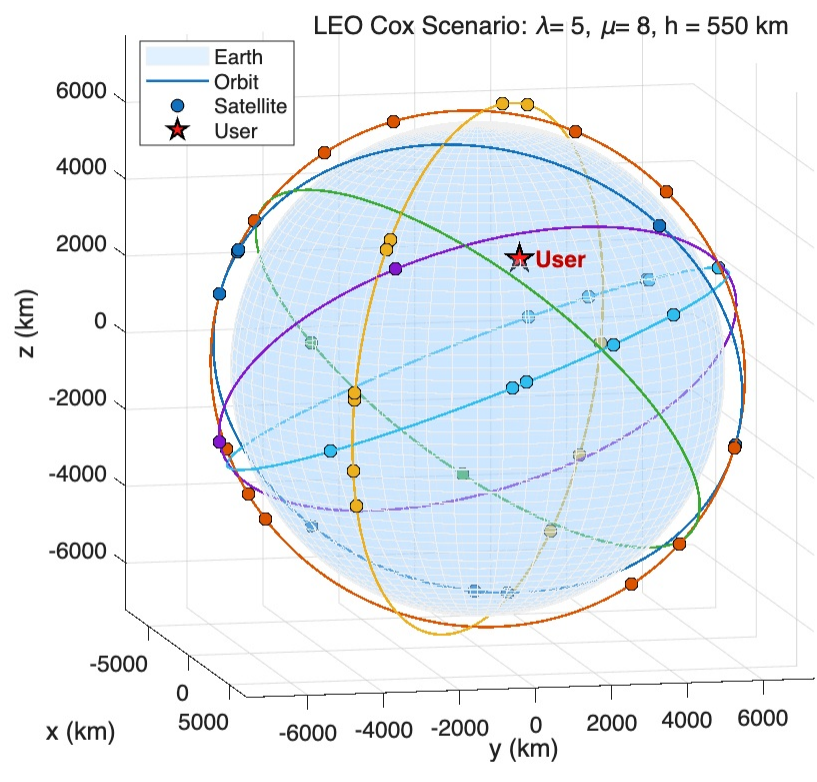}
	\caption{Schematic of the simulation scenario, including the Earth, a typical user, orbital planes, and satellites.}
	\label{fig6}
\end{figure}
\begin{figure}[!t]
	\centering
	\includegraphics[width=2.5in]{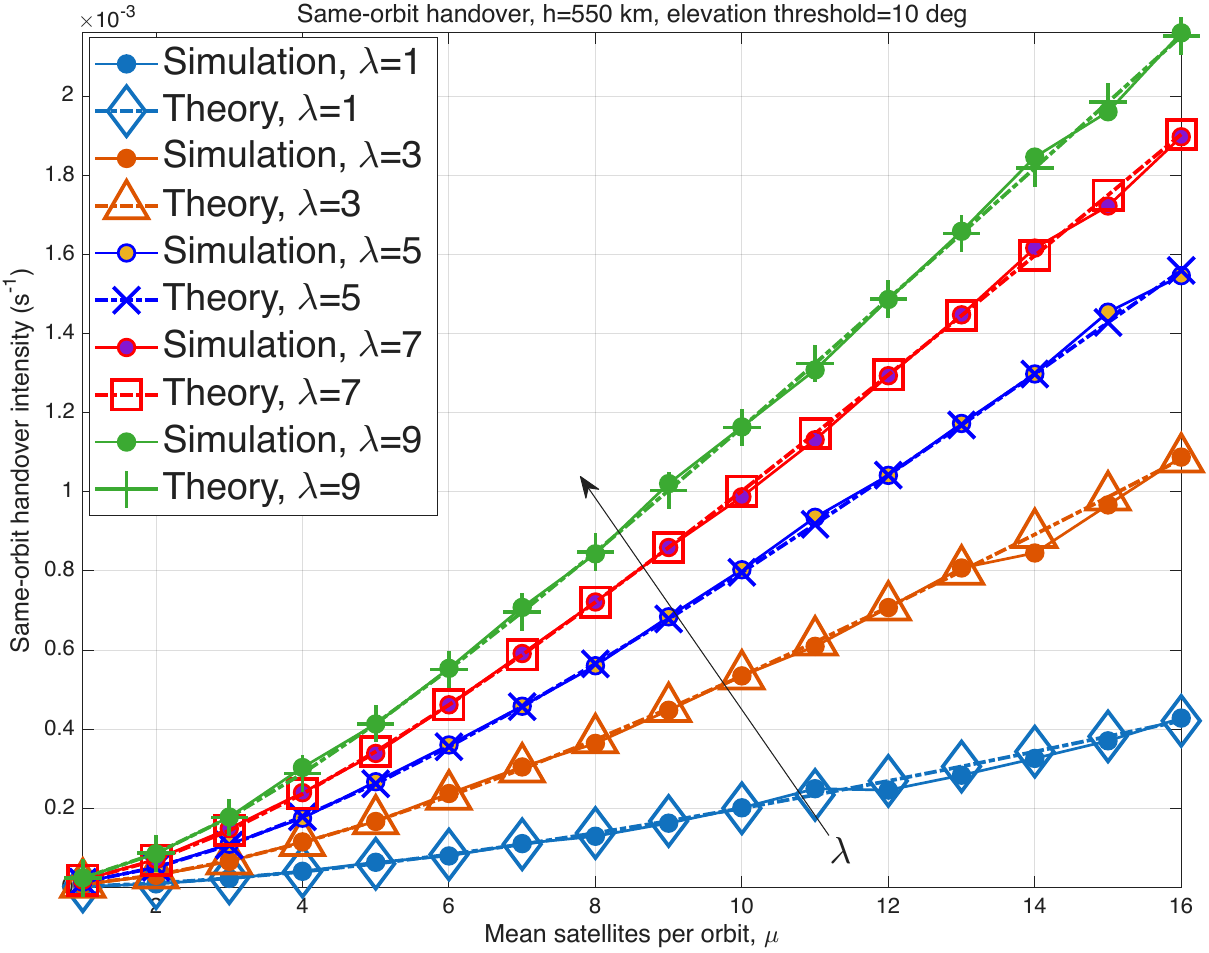}
	\caption{Same orbit handover intensity vs. satellite intensity per orbit $\mu$.}
	\label{fig7}
\end{figure}
\begin{figure}[!t]
	\centering
	\includegraphics[width=2.5in]{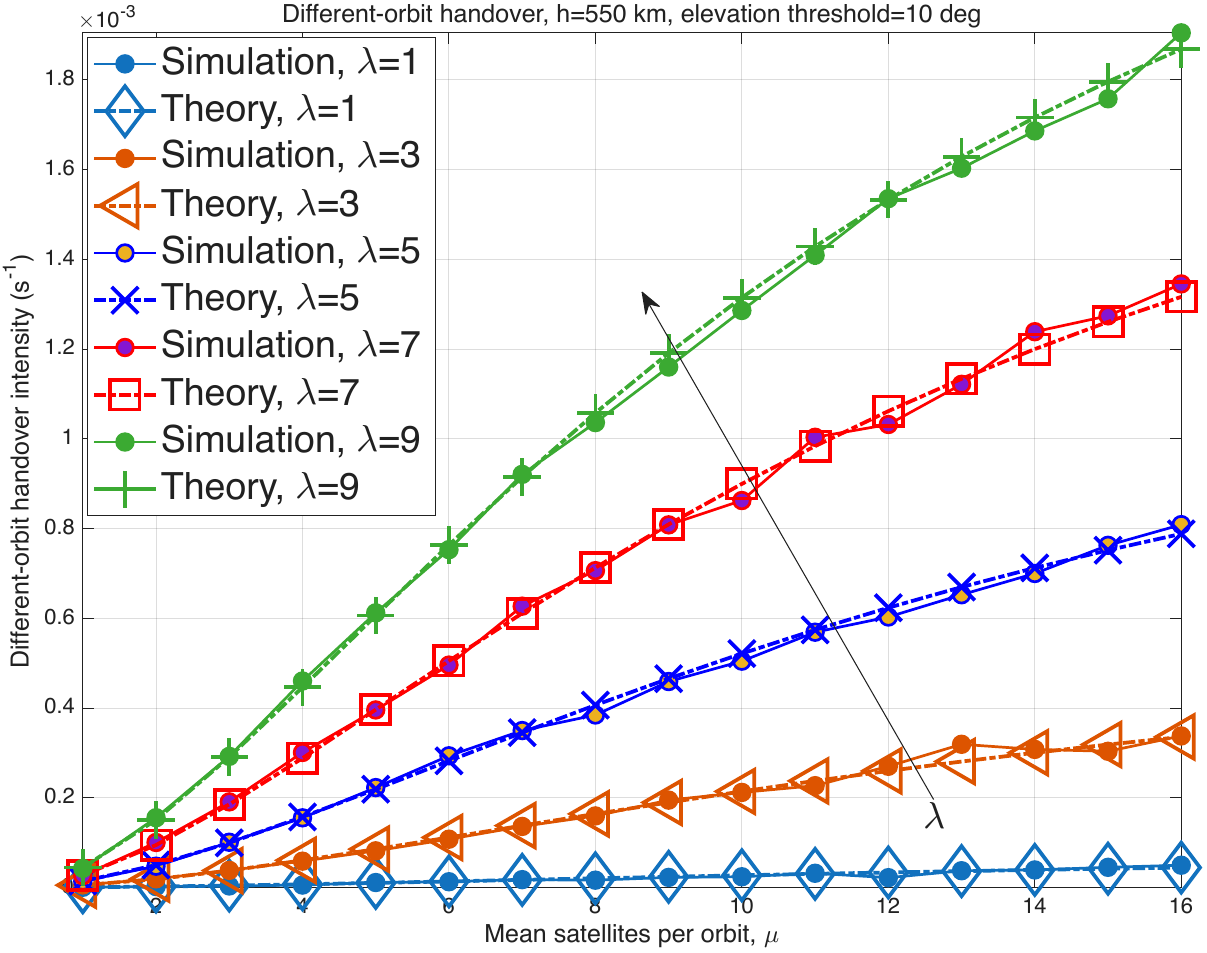}
	\caption{Different orbit handover intensity vs. satellite intensity per orbit  $\mu$.}
	\label{fig8}
\end{figure}
\begin{figure}[!t]
	\centering
	\includegraphics[width=2.5in]{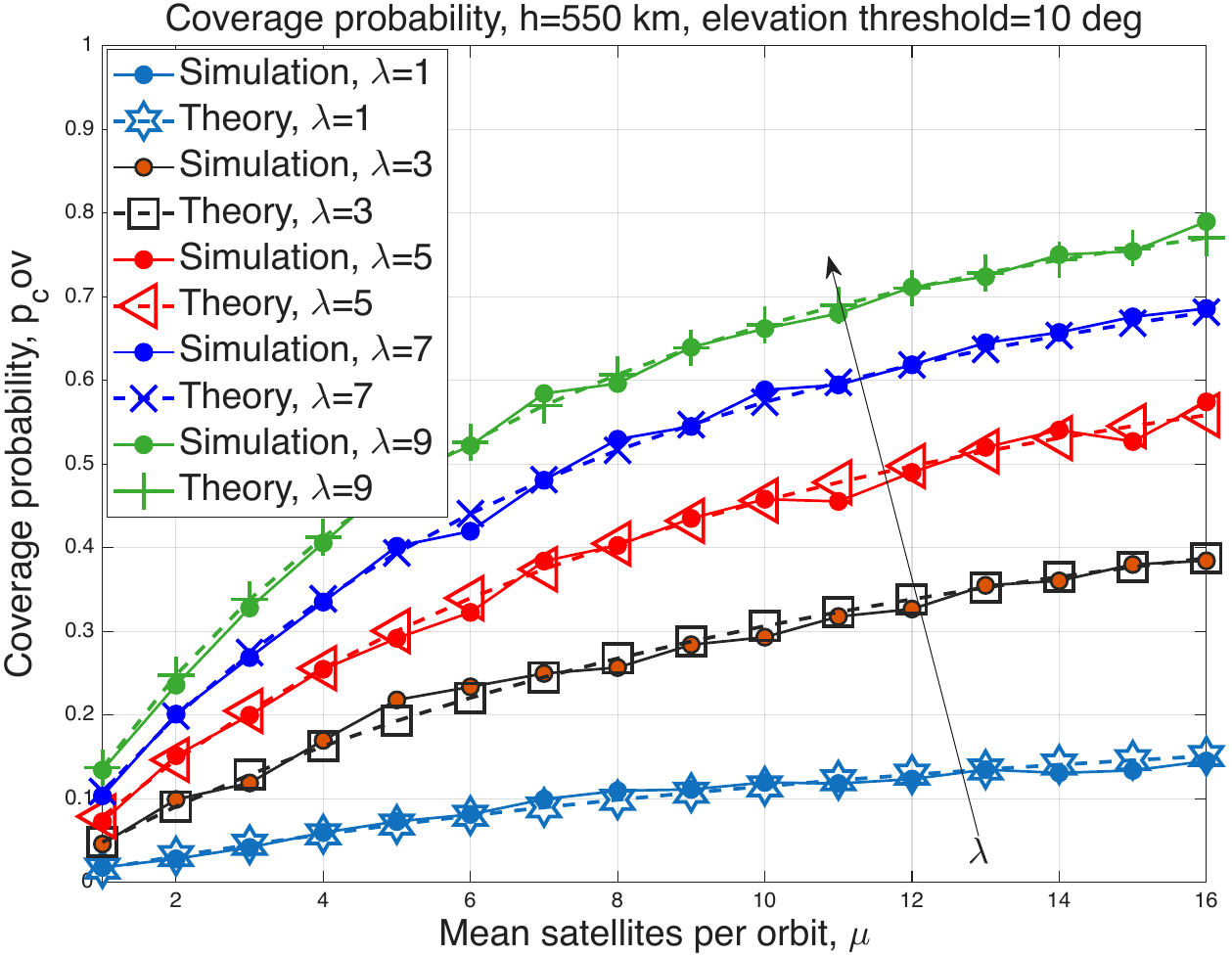}
	\caption{Coverage probability vs.satellite intensity per orbit $\mu$.}
	\label{fig9}
\end{figure}

\begin{figure}[!t]
	\centering
	\includegraphics[width=2.5in]{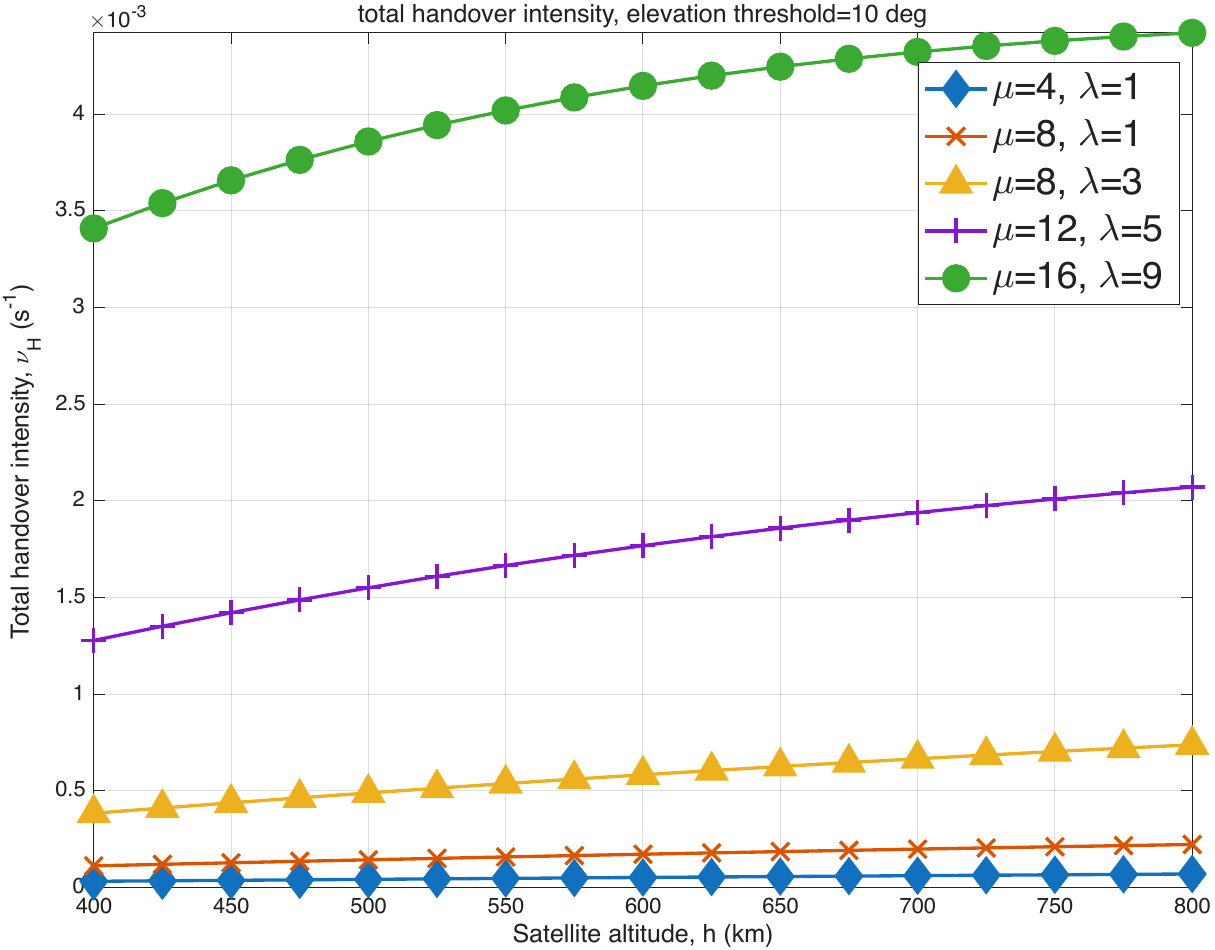}
	\caption{The impact of orbit height on handover intensity.}
	\label{fig10}
\end{figure}
The simulation scenario is shown in Fig.~\ref{fig6}, and the simulation results are obtained from the perspective of a typical user.
The solid curves in Figs.~\ref{fig7}, \ref{fig8}, and \ref{fig9} show the Monte Carlo simulation results. 
The dashed curves in Figs.~\ref{fig7}, \ref{fig8} show the numerical evaluation of the handover intensity for the same-orbit and different-orbit cases respectively. 
While the dashed curves in Fig.  \ref{fig9} show the numerical coverage probability. 
From these figures, it can be observed that the simulated and analytical curves match well over the tested parameter range.

As expected, the handover intensity increases as both the satellite and orbit intensities increase. As $\mu$ increases, each orbit is more likely to contain multiple satellites, which increases the likelihood of competing satellites. Similarly, increasing the orbit intensity $\lambda$ increases the number of independent parent orbits and therefore the likelihood of inter-orbit competition. The coverage probability in Fig.~\ref{fig8} also shows: increasing the satellite and orbit intensities does increase the likelihood that visible satellites exist.

Fig.~\ref{fig10} presents the impacts of orbit height on the total handover intensity over the selected $(\mu,\lambda)$ pair. 
For a low-density satellite constellation, increasing the orbit height does not significantly increase the handover intensity. 
In a higher-density deployment, such as $(\mu,\lambda)=(16,9)$, the handover intensity increases slightly with height.  
This counter-intuitive result is due to the fact that orbit height changes both the orbital angular speed and the visibility. The higher the orbit height, the more satellites become visible. When the visibility effect dominates the lower angular speed, the handover intensity increases. In a word, compared with satellite intensity, orbit height is not the main factor in reducing handover-related cost.

\section{Conclusion}
This paper proposes a new analytical framework for investigating the handover performance of dynamic LEO satellite networks in three-dimensional Euclidean space. The initial satellite positions are modeled as a classical Cox point process. From the perspective of a typical ground user, we devise a new static representation of the satellite geometry in a projection space, in which each orbit becomes a circle of random radius and each satellite becomes a fixed point. The motion of a satellite along its orbit is then characterized by the rotation of query vector. Consequently, the proximity between the typical user and a satellite can be characterized by the cosine distance, which is used to identify both the switching satellites and the locations of their competitors. Since the transformed radial Cox process preserves the shared orbit dependence, we exploit its reduced Palm distribution to derive the same-orbit and different-orbit handover intensities.

A toy example is used to validate the analytical results. Compared with previous literature, the proposed framework is both geometrically interpretable and analytically tractable under the nearest-visible-satellite association assumption. Simulations are conducted to verify the effectiveness of the analytical framework. As the intensities of satellites and orbits increase, the handover intensity increases proportionally. An increase in the constellation deployment height also slightly increases the handover intensity.

Future work may focus on the handover performance of multi-orbit constellations at different heights. The coverage performance and the persistence duration of the serving satellite will be investigated in more detail for different deployment scenarios.

\end{document}